\documentclass[letterpaper,twocolumn,showpacs,
prb,
,superscriptaddress,email
,showkeys%
]{revtex4-1}

\usepackage[pdftex]{graphicx}
\usepackage{amssymb}
\usepackage{mathrsfs}
\usepackage{amsmath,amsfonts,latexsym}
\usepackage{array,tabularx,color}
\usepackage{dcolumn}  
\usepackage[normalem]{ulem}
\usepackage{comment}
\usepackage{bm}
\usepackage{wasysym}
\usepackage{soul}

\begin{document}

\title{Quantum coherence in momentum space of light-matter condensates}

\author{C. Ant\'on}
\affiliation{Departamento de F\'isica de Materiales, Universidad Aut\'onoma de Madrid, Madrid 28049, Spain}
\affiliation{Instituto de Ciencia de Materiales ``Nicol\'as Cabrera'', Universidad Aut\'onoma de Madrid, Madrid 28049, Spain}

\author{G. Tosi}
\affiliation{Departamento de F\'isica de Materiales, Universidad Aut\'onoma de Madrid, Madrid 28049, Spain}

\author{M. D. Mart\'in}
\affiliation{Departamento de F\'isica de Materiales, Universidad Aut\'onoma de Madrid, Madrid 28049, Spain}
\affiliation{Instituto de Ciencia de Materiales ``Nicol\'as Cabrera'', Universidad Aut\'onoma de Madrid, Madrid 28049, Spain}

\author{Z. Hatzopoulos}
\affiliation{FORTH-IESL, P.O. Box 1385, 71110 Heraklion, Crete, Greece}
\affiliation{Department of Physics, University of Crete, 71003 Heraklion, Crete, Greece}

\author{G. Konstantinidis}
\affiliation{FORTH-IESL, P.O. Box 1385, 71110 Heraklion, Crete, Greece}

\author{P. S. Eldridge}
\affiliation{FORTH-IESL, P.O. Box 1385, 71110 Heraklion, Crete, Greece}

\author{P. G. Savvidis}
\affiliation{FORTH-IESL, P.O. Box 1385, 71110 Heraklion, Crete, Greece}
\affiliation{Department of Materials Science and Technology, University of Crete, 71003 Heraklion, Crete, Greece}

\author{C. Tejedor}
\affiliation{Departamento de F\'isica Te\'orica de la Materia Condensada, Universidad Aut\'onoma de Madrid, Madrid 28049, Spain}
\affiliation{Instituto de Ciencia de Materiales ``Nicol\'as Cabrera'', Universidad Aut\'onoma de Madrid, Madrid 28049, Spain}
\affiliation{Instituto de F\'isica de la Materia Condensada, Universidad Aut\'onoma de Madrid, Madrid 28049, Spain}

\author{L. Vi\~na}
\email{luis.vina@uam.es}
\affiliation{Departamento de F\'isica de Materiales, Universidad Aut\'onoma de Madrid, Madrid 28049, Spain}
\affiliation{Instituto de Ciencia de Materiales ``Nicol\'as Cabrera'', Universidad Aut\'onoma de Madrid, Madrid 28049, Spain}
\affiliation{Instituto de F\'isica de la Materia Condensada, Universidad Aut\'onoma de Madrid, Madrid 28049, Spain}

\date{\today}


\begin{abstract}

We show that the use of momentum-space optical interferometry, which avoids any spatial overlap between two parts of a macroscopic quantum state, presents a unique way to study coherence phenomena in polariton condensates. In this way, we address the longstanding question in quantum mechanics: \emph{``Do two components of a condensate, which have never seen each other, possess a definitive phase?"} [P. W. Anderson, \emph{Basic Notions of Condensed Matter Physics} (Benjamin, 1984)]. A positive answer to this question is experimentally obtained here for light-matter condensates, created under precise symmetry conditions, in semiconductor microcavities taking advantage of the direct relation between the angle of emission and the in-plane momentum of polaritons.

\end{abstract}

\maketitle

\section{Introduction}
\label{sec:intro}

Cold atoms and exciton-polaritons in semiconductor microcavities are systems where their capability to constitute Bose-Einstein condensates (BECs) has been demonstrated in recent years~\cite{Davis1995,kasprzak06:nature}. These BECs, due to their dual wave-particle nature, share many properties with classical waves as, for instance, interference phenomena~\cite{Andrews1997,Hall:1998bs,Esslinger:2000aa,Bloch:2005vn}, which are crucial to gain insight into their undulatory character~\cite{BornWolf2000,Ficek2005}. One of the main differences between atomic and polariton condensates resides in the particles lifetime: the finite lifetime of polaritons, in contrast with the infinite one of atoms, can be regarded as a complication. But making virtue of necessity, a short lifetime also implies a significant advantage: polaritons have a mixed exciton-photon character~\cite{kavokin13}, their lifetime being determined by the escape of their photonic component out of the cavity. These photons are easily measured either in real- (near field spectroscopy) or momentum-space (far field spectroscopy)~\cite{novotny2006}, rendering full information about the polariton BECs wave-function and, in particular, about its coherence~\cite{kasprzak06:nature}. Our goal is to profit from these measurements in momentum space to experimentally investigate something far from accessible in atomic condensates: the interference in momentum space produced by the correlation between two components of a condensate, which are, and have always been, spatially separated. Understanding coherence is important for a large number of disciplines spanning from classic optics to quantum information science and optical signal processing~\cite{Mandel1995,pryde2008}.

Pitaevskii and Stringari made a theoretical proposal to investigate experimentally these interference effects in momentum space via the measurement of their dynamic structure factor~\cite{Pitaevskii:1999fv}. In related experiments, coherence between two spatially separated atomic BECs has been indirectly obtained using stimulated light scattering~\cite{Saba:2005aa,Shin:2005aa}. In this work we perform a direct measurement of this correlation in polariton BECs, which moving in a symmetrical potential landscape, acquire a common relative phase, obtaining a positive answer to Anderson's question~\cite{anderson1984,Castin:1997aa,legget2006,Pitaevskii2003}, which opens new perspectives in the field of multi-component condensates.

\section{Experimental results and discussion}
\label{sec:exp}

\begin{figure}[htbp]
\begin{center}
\includegraphics[trim=0.0cm 0.3cm 5cm 1.5cm, clip=true,width=1\linewidth,angle=0]{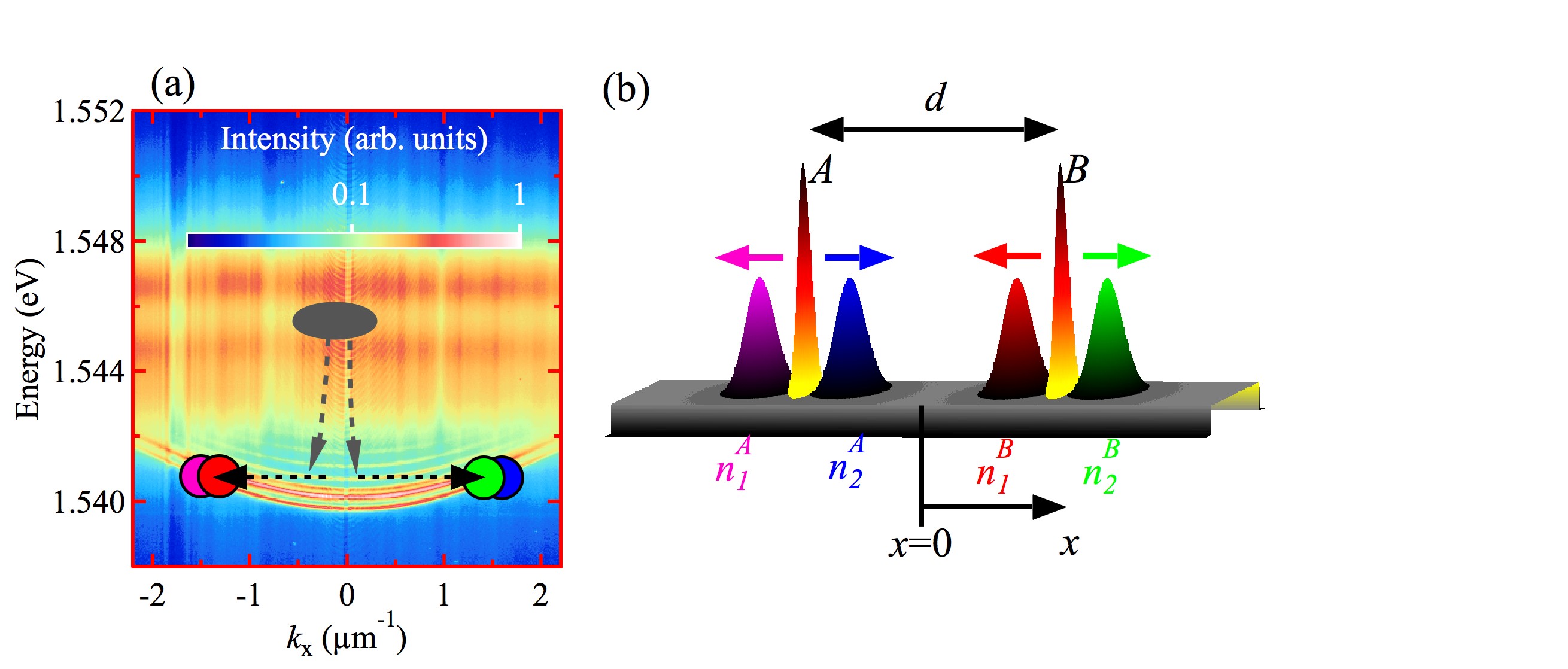}
\end{center}
\caption{(a) Sketch of the excitation and relaxation processes to form propagating polariton wave packets (\emph{WP}s) on a background showing the energy vs. $k_x$ emission obtained under non-resonant, low power excitation conditions. The grey ellipse depicts the excitation laser at 1.545 eV and $k_x \sim 0$. The dashed lines indicate the energy relaxation of excitons into polariton \emph{WP}s. Polariton \emph{WP}s, propagating with  $k_x \approx \pm1.6 $ $\mu$m$^{-1}$ (slightly displaced for the sake of clarity), are depicted with circles, coded in colors explained in (b). The emission intensity is coded in a logarithmic, false color scale. (b) Sketch in real space of the experimental configuration. A laser beam is split into two arms, \emph{A} and \emph{B}, distanced by $d$. They create four propagating polariton \emph{WP}s, coded in different colors, $n_{1,2}^A$ (magenta, blue) and $n_{1,2}^B$ (red, green) moving along the $x$ axis of a microcavity ridge in the direction depicted by the arrows.}
\label{fig:fig1}
\end{figure}

We confront this task in a quasi one-dimensional (1D) system made of a high-quality AlGaAs-based microcavity, where $20 \times 300$ $\mu$m$^2$ ridges have been sculpted. The sample, kept at 10 K, is excited with 2 ps-long light pulses from a Ti:Al$_2$O$_3$ laser. In order to create polaritons in two separated spatial regions, the laser beam is split in two, named \emph{A} and \emph{B}, impinging simultaneously at positions distanced by $d_{AB}=70$ $\mu$m. Additional experimental details are described in the Supplementary information~\cite{supple}. A crucial issue when optically creating polaritons is the excess energy of the excitation laser. There are two well explored alternatives: non-resonant excitation at very high energies~\cite{kasprzak06:nature} and strictly resonant excitation~\cite{amo2009_b}. The latter situation generally produces macroscopic polariton states with a phase inherited from that of the laser, unless special care is taken in the experiments~\cite{Amo:2011qf}. The former case is appropriate to avoid phase heritage, but it does not provide the momentum distribution, shown below, required for our experiments. In order to avoid these difficulties, we opt for a different alternative, depicted in Fig.~\ref{fig:fig1}(a): the laser beams excite the sample at the energy of bare excitons and $k_x\sim0$. The broad bands between 1.542 and 1.548 eV corresponds to excitonic emission bands; the sub-bands below 1.542 eV are the confined lower polariton branches. After energy relaxation, polariton condensates are created in a process that involves a non-reversible dressing of the excitons and therefore an erasure of the laser phase~\cite{supple}.
Above a given pump intensity threshold, polaritons with $k_x\sim0$ evolve towards two states with momenta $\pm k_x$ (Fig.~\ref{fig:fig1}(a)). As sketched in Fig~\ref{fig:fig1}(b), this procedure results in the formation of four propagating polariton wave packets (\emph{WP}s). We label the macroscopic state of the \emph{WP}s as $\psi_1^A$, $\psi_2^A$, $\psi_1^B$, $\psi_2^B$, where the superscript refers to the excitation beam, the subscript $1$($2$) is for \emph{WP}s initially moving to the left (right), i.e. with $k_x < 0$ ($k_x > 0$). The direction of propagation is determined by the presence of local effective-barrier potentials ($V_A$ and $V_B$), associated to a blue-shifted dispersion relation, coming from carrier-carrier repulsive interactions~\cite{Wertz:2010ys}. The densities of the polariton \emph{WP}s are given by $n^{A,B}_j=\left|\psi^{A,B}_j\right|^2,~j=1,2$.

\emph{WP}s created by \emph{A} have never been together with those generated by \emph{B}, as sketched in Fig.~\ref{fig:fig1}(b). However, \emph{WP}s with the same subscript $j$ are in the same quantum state~\cite{note2_kx}. Using the capability of measuring directly in momentum space, a unique condition only achievable in light-matter condensates, we can assess whether or not \emph{WP}s $\psi_1^A$ and $\psi_1^B$ (or $\psi_2^A$ and $\psi_2^B$) are correlated to each other, being components of the same condensate. The two \emph{WP}s propagating to the left are described by a common macroscopic order parameter
\begin{align}
\Psi_1^{coh}\left(x\right)=\psi_1^{A}\left(x\right)+e^{i \phi}\psi_1^{B}\left(x\right),
\label{eq:eq0}
\end{align}
while those propagating to the right are described by
\begin{align}
\Psi_2^{coh}\left(x\right)=e^{i \phi}\psi_2^{A}\left(x\right)+\psi_2^{B}\left(x\right).
\label{eq:eq1}
\end{align}
The phases are chosen to have inversion symmetry with respect to $x=0$, because in our experiments we tune the intensities of the two lasers in order to get a symmetrical potential $V\left(x\right)=V\left(-x\right)$.  In that respect, our condensates are related to each other through the symmetry of the excitation process.

Furthermore, our potential landscape renders an equal motion for $\psi_j^A$ and $\psi_j^B$, i.e. equal momenta $|\left(k_x\right)_j^A|=|\left(k_x\right)_j^B|=k_x$. These are precisely the suitable conditions to observe coherence between two components spatially separated by $d$, i.e. $\psi_j^A\left(x-d/2\right)=\psi_j^B\left(x+d/2\right)=\psi_0\left(x\right)$, of a given condensate $\Psi_j^{coh}$. This coherence can be observed in $\mathbf{k}$-space as we discuss now.

For the sake of clarity, we focus in the following discussion only on the left-propagating \emph{WP}s. The corresponding order parameter in \textbf{k}-space can be written as:
\begin{align}
\Psi^{coh}_1\left(k_x\right)=\psi^{A}_1\left(k_x\right)+e^{i\phi}\psi^{B}_1\left(k_x\right)=\notag\\
&\hspace{-40mm}e^{-i k_x d/2} \psi_0\left(k_x\right)+e^{i\left(\phi+k_x d/2\right)} \psi_0\left(k_x\right)\label{eq:eq2}
\end{align}
with $\psi_0\left(k_x\right)$ being the Fourier transform of $\psi_0\left(x\right)$~\cite{Pitaevskii:1999fv}. This yields a momentum distribution
\begin{align}
n^{coh}_1\left(k_x\right)=\left|\Psi^{coh}_1\left(k_x\right)\right|^2=2\left[1+cos\left(k_x d + \phi\right)\right]\left|\psi_0\left(k_x\right)\right|^2.
\label{eq:eq3}
\end{align}
The coherence between the two components produces interference fringes with a period
\begin{align}
\Delta k_x=2\pi/d.
\label{eq:eq4}
\end{align}
Our aim is to observe the existence of interferences in $\mathbf{k}$-space coming from this macroscopic two-component condensate. Far-field detection allows the direct measurement of momentum distributions, i.e. it gives a direct determination of the existence, and the period, of these interference fringes. It must be taken also into account that the measured total polariton density is formed by a condensed population, $n^{coh}$, coexisting with a thermal one~\cite{Valle:2009aa}, therefore the interference patterns visibility, $\nu$, is lower than 1 (see Supplemental Material~\cite{supple}).

\begin{figure*}[htbp]
\begin{center}
\includegraphics[trim=0.4cm 0.3cm 0.4cm 0.4cm, clip=true,width=0.75\linewidth,angle=0]{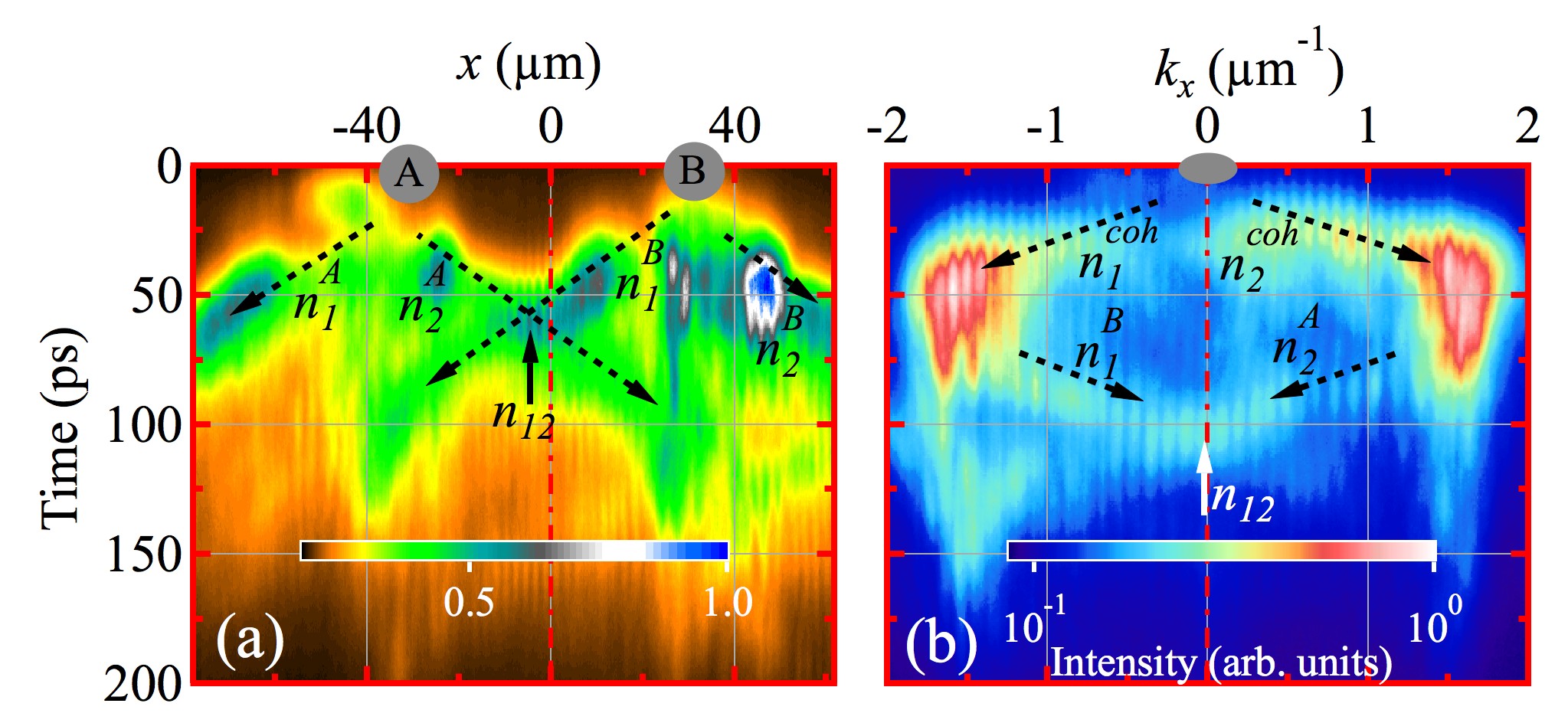}
\end{center}
\caption{(a) Emission in real space, along the $x$ axis of the ridge, versus time. Gray circles at $x=\pm35$ $\mu$m indicate the spatial location of the \emph{A} and \emph{B} laser beams; the trajectories of the four \emph{WP}s, $n_1^A$, $n_2^A$, $n_1^B$ and $n_2^B$, are indicated by the dashed arrows. (b) Momentum space emission, along $k_x$, versus time. The grey circle indicates that the laser beams, \emph{A} and \emph{B}, excite the ridge at $k_x \sim 0$. The dashed, black arrows indicate the acceleration of the condensates $n_1^{coh}$ and $n_2^{coh}$, as well as the deceleration of the \emph{WP}s $n_1^{B}$ and $n_2^{A}$. Intensity is coded in a normalized, logarithmic false color scale.}
\label{fig:fig2}
\end{figure*}

Our most important result is shown in Fig.~\ref{fig:fig2}(b): we indeed observe the interference fringes in $\mathbf{k}$-space, described by Eq.~\ref{eq:eq3}, directly in the polariton emission. This certifies the correctness of our hypothesis that each couple of \emph{WP}s ($\psi_j^A$, $\psi_j^B$) constitutes a two component condensate. Figure~\ref{fig:fig2}(a) shows the actual evolution in time of the four \emph{WP}s schematically depicted in Fig.~\ref{fig:fig1}(a): our results clearly demonstrate that the distance $d$ between the two components of each condensate remains constant with time during the first $\sim70$ ps ($d= d_{AB}$), as evidenced by the dashed parallel arrows. Figure~\ref{fig:fig2}(a) contains also interesting real-space interferences when \emph{WP}s $\psi_2^A$  and $\psi_1^B$ overlap in real space at 66 ps that we shall discuss in more detail below. A peculiarity of our experiments is that we observe the dynamics of the coherence; this allows us to determine that the two components of the condensate are phase locked since there is not any drift in the interference patterns.

As readily seen in Fig.~\ref{fig:fig2}(b), an initial acceleration of the four \emph{WP}s, from rest, $k_x = 0$, to $k_x=\pm1.6$ $\mu$m$^{-1}$ during the first 40 ps, is followed by a uniform motion taking place from 40 ps to 70 ps. The interference pattern of each condensate is observed until $\sim75$ ps, instant at which $\psi_1^A$ and $\psi_2^B$ disappear from the sample region imaged in the experiments. Then \emph{WP}s $\psi_1^B$ and $\psi_2^A$  are progressively slowed by the presence of the barriers at the excitation spots ($V_A$/$V_B$ halts $\psi_1^B$/$\psi_2^A$). When these two \emph{WP}s, which are the components of two different condensates $\Psi_1^{coh}$ and $\Psi_2^{coh}$, are stopped (at $\sim100$ ps) another interference appears in $\mathbf{k}$-space, but now at $k_x = 0$ as it corresponds to \emph{WP}s at rest. This means that these two condensates also interfere with each other, being remarkable that $\Psi_1^{coh}$ and $\Psi_2^{coh}$ still preserve some kind of mutual coherence, supporting the functional form of Eqs.~(\ref{eq:eq0}) and (\ref{eq:eq1}). For longer times, the two \emph{WP}s move again, as can be observed in Figs.~\ref{fig:fig2}(a,b), becoming more difficult to track their trajectories.

\begin{figure*}[htbp]
\begin{center}
\includegraphics[trim=0.4cm 1.1cm 0.4cm 0.3cm, clip=true,width=1\linewidth,angle=0]{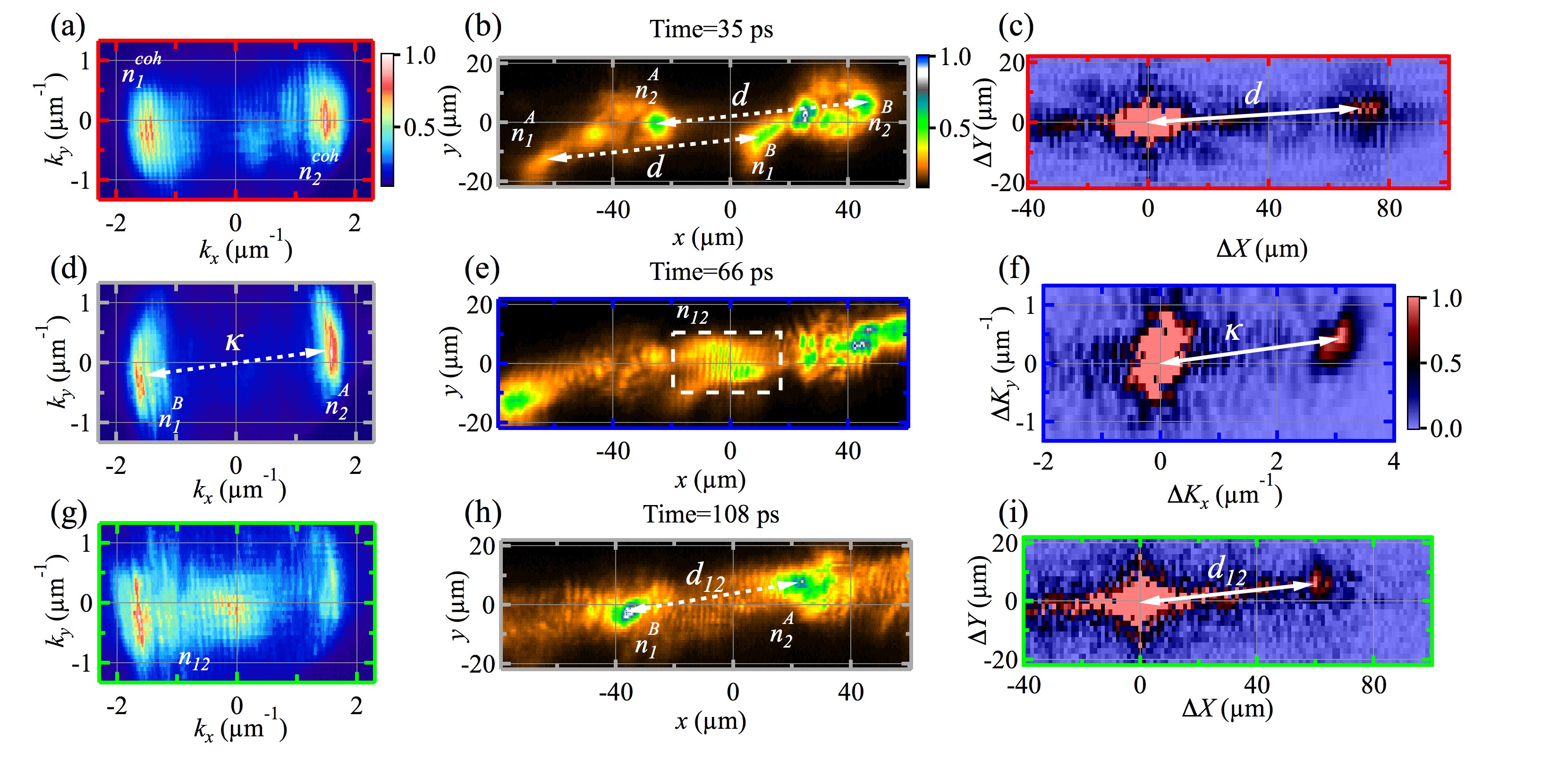}
\end{center}
\caption{(a) Momentum distribution $n\left(\mathbf{k}\right)$, at 35 ps after the excitation, showing the condensates $n_1^{coh}$/$n_2^{coh}$ at $k_x=\mp 1.6$ $\mu$m$^{-1}$, respectively. (b) Corresponding $n\left(\mathbf{r}\right)$ distribution showing \emph{WP}s $n_1^A$, $n_2^A$, $n_1^B$ and $n_2^B$.  (c) Fourier transform of $n\left(\mathbf{k}\right)$, obtaining a frequency at $\Delta X=d=70$ $\mu$m. (d) Momentum distribution $n\left(\mathbf{k}\right)$ at 66 ps showing  $n_1^B$ and $n_2^A$ at $k_x=\mp 1.6$ $\mu$m$^{-1}$, respectively.  (e) Real space distribution $n\left(\mathbf{r}\right)$ showing the interferences of $n_{12}$ at $x= 0$, created by the overlapping in real space of $\psi_1^B$ and $\psi_2^A$. White dashed rectangle marks the region of interest where the interference occurs. (f) Fourier transform restricted to the region of interest in $n\left(\mathbf{r}\right)$, showing a frequency at $\Delta K_x = \kappa = 3.2$ $\mu$m$^{-1}$. (g) Momentum distribution $n\left(\mathbf{k}\right)$ at 108 ps, showing the interferences $n_{12}$ at $k_x \sim 0$. (h) Corresponding $n\left(\mathbf{r}\right)$ distribution showing $n_1^B$ and $n_2^A$.  (i) Fourier transform of $n\left(\mathbf{k}\right)$, obtaining a frequency at $\Delta X = d_{12} = 60$ $\mu$m. Intensities in the false color scales for momentum, real and Fourier spaces are normalized to unity. The tilt in all panels originates from the orientation of the ridge with respect to the entrance slit of the spectrometer. The white dashed arrows mark the distances in real- and momentum-space between \emph{WP}s. The full arrows show these distances in the corresponding Fourier transform. Supplementary Video S1/S2 shows the time evolution of the emission in real/momentum space~\cite{supple}.}
\label{fig:fig3}
\end{figure*}

Note that our measurements are performed averaging over millions of shots of the pulsed laser, therefore if $\phi$ were a phase determined by the projection involved in the measurement process~\cite{Castin:1997aa,legget2006}, it would take a random value in each realization. Then, averaging over all the possible results, the interference pattern would not be observed. However, as a consequence of the symmetry $V(x)=V(-x)$ of the potential, the whole state of the four \emph{WP}s, $\Psi$, is symmetric, both in real- and momentum-space. The continuity in \textbf{k}-space of the wave-function ($\Psi (k_x)$) and of its derivative ($\partial \Psi (k_x)/\partial k_x$) sets the relative phase $\phi$ and makes the experimental realizations contribute constructively to the observed interference patterns. In other words, the spatial symmetry involved in the buildup of the condensates determines the relative phase $\phi$. In this sense, they are not independent from each other although they have never before coincided in real space.

Further insight into the quantum coherence is obtained by analyzing in detail the interferences occurring in momentum- and real-space. Accordingly, we present in Fig.~\ref{fig:fig3} two-dimensional maps of the polariton emission at three consecutive, relevant times~\cite{note1_nat}. We focus on the correspondence between the period of the interference patterns in each space (real and momentum) and the separation between the \emph{WP}s in the complementary space. Figure~\ref{fig:fig3}(a) shows the momentum distribution $n\left(k_x,k_y\right)$, 35 ps after the impinging of the laser beams on the sample. The coherence of each $\Psi_j^{coh}$ is observed by the conspicuous interference patterns, $n_j^{coh}$, centered at $k_x=\pm1.6$ $\mu$m$^{-1}$. In both cases, the fringes period amounts to $\Delta k_x=0.088(5)$ $\mu$m$^{-1}$ that, according to Eq.~\ref{eq:eq4}, should correspond to a distance between \emph{WP}s of $d= 71(4)$ $\mu$m. This is in good agreement with the experimental distance seen in Fig.~\ref{fig:fig3}(b): the two components of each condensate, $n_j^A$ and  $n_j^B$, are separated by $d\simeq 70$ $\mu$m (see dashed arrows). Our findings are further supported by the Fourier transform map of $n\left(k_x,k_y\right)$ shown in Fig.~\ref{fig:fig3}(c): a well-defined Fourier component at $\Delta X=d=70$ $\mu$m is obtained, in accordance with the separation directly observed in real space.

Coherence in real space have been profusely studied in cold atoms~\cite{Andrews1997,Esslinger:2000aa,Hodgman:2011aa}, excitons~\cite{Snoke:2002aa,High2012} and polariton condensates~\cite{kasprzak06:nature,Balili2007,Roumpos2012,Manni2012,Rahimi-Iman:2012ij,Spano2012}. Our experiments also show interferences in real space between two condensates, similar to those reported in atomic BECs~\cite{Andrews1997,Esslinger:2000aa}. This is shown in Fig.~\ref{fig:fig3}(e) at 66 ps when \emph{WP}s $\psi_2^A$ and $\psi_1^B$ meet each other at $x\sim 0$. The appearance of interference fringes in real space, $n_{12}$, signals unambiguously to coherence between these two \emph{WP}s. Since real and momentum spaces are reciprocal to each other, equivalent results for the interference patterns are expected. The complementary expression in real space to Eq.~\ref{eq:eq4} reads now $\Delta x=2\pi/\kappa$,
where $\Delta x$ is the period of the fringes and $\kappa$ the difference in momentum of the propagating \emph{WP}s. The experimental period of the fringes, seen in the dashed-rectangle area in Fig.~\ref{fig:fig3}(e), $\Delta x=1.99(17)$ $\mu$m, should correspond to $\kappa=\left(k_x\right)_2^A-\left(k_x\right)_1^B=3.2(2)$ $\mu$m$^{-1}$. This is again borne out by our results, as shown in Fig.~\ref{fig:fig3}(d), where the emission in $\mathbf{k}$-space shows clearly that \emph{WP}s $\psi_2^A$ and $\psi_1^B$ are counter-propagating with $k_x=\pm1.6$ $\mu$m$^{-1}$, respectively. Figure~\ref{fig:fig3}(f) shows the Fourier transform of $n_{12}$ in the region enclosed by the rectangle in Fig.~\ref{fig:fig3}(e). It reveals a strong $\Delta K_x$ Fourier component at 3.1 $\mu$m$^{-1}$, in full agreement with the value of $\kappa$ displayed in Fig.~\ref{fig:fig3}(d). Let us also emphasize that \emph{WP}s first meet in real space at 66 ps, while interferences in momentum space are seen as early as $\sim10$ ps demonstrating that the phase locking occurs before the \emph{WP}s spatially overlap.

The third result that we present corresponds to the arrival at 108 ps of $\psi_2^A$ and $\psi_1^B$ to the excitation regions \emph{B} and \emph{A}, respectively. Here, they run into the hills of the photogenerated potentials $V_B$ and $V_A$ that elastically convert their kinetic energy into potential energy~\cite{Anton:2013aa}. They slow down, halting, providing a new separation between \emph{WP}s $n_2^A$ and $n_1^B$, $d_{12} \sim 60$ $\mu$m (see Fig.~\ref{fig:fig3}(h)). Their emission in momentum space, arising from $k_x\sim 0$, evidences an interference pattern with $\Delta k_x=0.108(5)$ $\mu$m$^{-1}$ ($n_{12}$, see Fig.~\ref{fig:fig3}(g)). Once again, Eq.~\ref{eq:eq4} predicts a separation $d_{12}=60(4)$ $\mu$m between $n_2^A$ and $n_1^B$, as observed in the experiments. For completeness, we also show in Fig.~\ref{fig:fig3}(i) the Fourier transform map of the density that exhibits an emerging component at $\Delta X=d_{12}=60$ $\mu$m. Further insight into this scaling behavior, relating distances in real space between \emph{WP}s with the fringes period in momentum space, is presented in the Supplementary information~\cite{supple}.

\section{Conclusions}
\label{sec:conclu}

In summary, the convenience of monitoring the evolution of exciton-polaritons in semiconductor microcavites, through the detection of emitted light, makes this system an ideal platform to study quantum coherence properties in real- as well as in momentum-space. Profiting from this fact, we have demonstrated the existence of quantum remote coherence between spatially separated polariton condensates whose phase is determined by the symmetry of the excitation conditions and therefore is constant in each realization of our multi-shot experiments. This issue is related to the superposition principle in quantum mechanics and it is crucial to understand how mutual coherence is acquired.

\section{Acknowledgements}
\label{sec:acknow}

We thank D. Steel and J.J. Baumberg for a critical reading of the manuscript. C.A. acknowledge financial support from a Spanish FPU scholarship. P.G.S. acknowledges Greek GSRT program ``ARISTEIA" (1978) for financial support. The work was partially supported by the Spanish MEC MAT2011-22997, CAM (S-2009/ESP-1503) and FP7 ITN's ``Clermont4" (235114), ``Spin-optronics" (237252) and ``INDEX" (289968) projects.





\begin{thebibliography}{10}
\bibitem{note1_nat}
Since our samples are not strictly one-dimensional, we provide full images of the emission in two-dimensional spaces.
\end{thebibliography}


\end{document}